\documentstyle[12pt,epsfig,graphicx]{article}
\def\beq{\begin{equation}}  
\def\eeq{\end{equation}}  
\def\bea{\begin{eqnarray}}   
\def\eea{\end{eqnarray}}   
\def\bq{\begin{quote}}   
\def\eq{\end{quote}}   
\def\bi{\begin{itemize}}   
\def\ei{\end{itemize}}   
\def\beqa{\begin{eqnarray}}   
\def\eeqa{\end{eqnarray}}   
\def\be{\begin{enumerate}}   
\def\ee{\end{enumerate}}   
\def\beq{\begin{equation}}   
\def\eeq{\end{equation}}   
\def\bi{\begin{itemize}} 
\def\ei{\end{itemize}}

\def\pa{\partial}

\def\cp{{\cal P}}   
\def\cl{{\cal L}}  
  
\def\cq{{\cal Q}} 
\def\cR{{\cal R}} 
\def\cS{{\cal S}}

\def\r2{\sqrt{2}}   
\def\ra{\rightarrow} 
\def\bi{\begin{itemize}}   
\def\ei{\end{itemize}}

\def\ov{\overline}   
\def\nn{\nonumber \\}

\def\ca{{\cal A}}

\begin{document} 
%%%%%%%%%%%%%%%%%%%Strona Tytulowa%%%%%%%%%%%%%%%%%%%%%%%%%%%%%%%%%% 
\pagestyle{empty} 
\begin{flushright}   CERN-TH/2001-198
\end{flushright} 
\vskip 2cm 
\begin{center} 
{\Huge Non-BPS Branes of  Supersymmetric\\ Brane Worlds} 
\vspace*{5mm} \vspace*{1cm}  
\end{center} 
\vspace*{5mm} \noindent 
\vskip 0.5cm 
\centerline{\bf Philippe Brax${}^{1,2}$, Adam Falkowski${}^3$ and Zygmunt Lalak${}^{1,3}$} 
\vskip 1cm 
\centerline{\em ${}^{1}$Theory Division, CERN} 
\centerline{\em CH-1211 Geneva 23, Switzerland} 
\vskip 0.3cm 
\centerline{\em ${}^{2}$ Service de Physique Th\'eorique} 
\centerline{\em CEA-Saclay F-91191 Gif/Yvette, France} 
\vskip 0.3cm 
\centerline{\em ${}^{3}$Institute of Theoretical Physics} 
\centerline{\em University of Warsaw, Poland} 
\vskip 2cm 
 
\centerline{\bf Abstract} 
\noindent We consider five-dimensional brane worlds
with $N=2$ gauged  
supergravity in the bulk coupled supersymmetrically to two boundary branes 
at the fixed points of a $Z_2$ symmetry. We analyse two mechanisms
that break supersymmetry either by choosing flipped fermionic 
boundary conditions on the boundary branes or by modifying the gravitino 
variation 
to include both $Z_2$-odd and $Z_2$-even operators. In all cases we 
find the corresponding background. Including an even part in the gravitino 
variation leads to tilted branes.  Choosing the flipped boundary
conditions leads to $AdS_4$ branes and stabilized radion in the detuned
case,
when the expectation value of the even variation is nonzero.
Another solution has the interpretation of moving $AdS_4$ branes separated
by a horizon. The solution with moving branes
separated by a horizon can be extended to the tuned case. In the presence
of a horizon,
temperature mediation communicates supersymmetry breakdown
to the branes.

\vskip .3cm 
%%%%%%%%%%%%%%%%%%%%%%%%%%%%%%%%%%%%%%%%%%%%%%%%%%%%% 
\newpage 
\pagestyle{plain}  
\section{Introduction} 
It has been realized recently that there exist locally supersymmetric  
theories in five dimensions that include nontrivial physics localized on  
four dimensional branes \cite{bagger,gp,flp,flp2,kallosh}. The brane sectors may contain arbitrary  
four dimensional gauge theories as well as localized interactions between  
bulk fields. In the bulk one has a gauged 5d supergravity, which can be  
further coupled to five dimensional gauge sectors. Such five dimensional  
models with branes are likely to lead to interesting extensions of  
the Standard Model, pertaining to novel approaches to the hierarchy problem. 
The class of brane--bulk supergravities that has been studied so far  
in some detail, \cite{flp3}, is the one that allows for BPS solutions in five dimensions, 
thus resulting in a class of four-dimensional supergravities.  
There exists however a possibility, that Lagrangians that are locally
supersymmetric in the whole 5d space, including branes, do not allow
for a  
supersymmetric, BPS, vacuum. Such scenarios, where five-dimensional  
supersymmetry is  
broken down to nothing by the vacuum configurations were considered in the  
local case by Fabinger and Horava, \cite{Fabinger:2000jd}, and their 
phenomenology at the level of  
{\em globally} (non)supersymmetric models has been studied by Barbieri et.  
al. and in the subsequent papers \cite{Barbieri:2001vh} (see also \cite{Gherghetta:2001kr} and \cite{Antoniadis:1990ew,Antoniadis:1994jp} for related discussion). However, eventually 
warped gravity and moduli fields have to be consistently included into the game. In this note we make some steps in  
this direction. We study explicitly gauged supergravity coupled supersymmetrically to four-dimensional  branes. However, we generalize the construction of \cite{flp,flp2} in two ways. Firstly, we extend supersymmetry transformations,  
and the prepotential, by $Z_2$-even terms. These terms introduce  
supersymmetric detuning  
between brane tensions and bulk potential energy. Secondly, we allow  
the $Z_2$-projections to be different on different branes.  
These new ingredients give rise to interesting new families of vacuum  
solutions, which are in fact examples of non-BPS branes.  
We give the explicit form of vacuum configurations, and discuss their  
properties.  
 
The supergravities we consider in this paper form a background for  
interesting particle physics. After inserting supersymmetric gauge sectors  
on the branes, supersymmetry breakdown will be transmitted to the  
chiral models  
living there, but the supersymmetric pattern of couplings will be preserved.  
In the case of a charged bulk sector, the mass splits in the multiplets will 
be caused directly by different boundary conditions for bosons and fermions.  
The `flipped' models are five--dimensional representatives of $Dp$--$D\bar{p}$ 
brane--antibrane systems. 
The work we present here should help understanding the low--energy physics of 
these interesting theories.  
 
\section{Supersymmetry Breaking}     
Let us recall the basic features of pure 5d N=2 gauged supergravity. The gravity multiplet $(e_\alpha^m, \psi_\alpha^A, \ca_\alpha)$ consists of  the metric (here we use a vielbein), a pair of symplectic Majorana gravitini, and a vector field called the graviphoton. There is a global SU(2) R-symmetry which rotates the two supercharges into each other. Making use of the graviphoton we can gauge a U(1) subgroup of the R-symmetry group. Such gauging can be described by an SU(2) algebra valued matrix $\cp=\vec{P} \cdot  i\vec{\sigma}$ (prepotential). We do not give the complete form of the  action and supersymmetry transformation laws in gauged supergravity (see \cite{agata} for details), but only the relevant 
terms. The gravitino transformation law gets the following correction due to gauging (we use the normalization of \cite{ovrut}) 
\beq 
\label{susygrav} 
\delta \psi_\alpha^A =  
-i\frac{\r2}{3}\gamma_\alpha g\cp^A_B \epsilon^B    
\eeq        
where $g$ is the U(1) gauge charge. Gauging introduces also the potential term into the action 
\beq 
V =\frac{8}{3} g^2 Tr (\cp^2). 
\eeq    
Without bulk matter fields the prepotential is just a constant matrix so the potential term corresponds to a (negative) cosmological constant. 
 
The distinguishing feature of the Randall-Sundrum scenario is that the fifth dimension is an orbifold $S_1/Z_2$. It is equivalent (and technically easier) to work on a smooth circle $S_1$ with the fifth coordinate ranging from $-\pi\rho$ to $\pi\rho$ and impose $Z_2$ symmetry on the fields of our Lagrangian.     
The $Z_2$ symmetry acts by $x^5 \ra -x^5$. Under its action the bosonic fields $g_{\mu\nu}$, $g_{55}$ and  $\ca_5$ are even, while $g_{\mu 5}$ and $\ca_\mu$ are odd. The $Z_2$ action on the gravitino is defined as follows 
\bea 
\label{z2} 
\psi^A_\mu(-x^5)= \gamma_5 \cq^A_B \psi_\mu^B(x^5) & \psi^A_5(-x^5)= -\gamma_5 \cq^A_B \psi_5^B(x^5) 
\eea 
where $\cq=\vec{Q} \cdot  \vec{\sigma}$ and $\vec{Q}^2=1$. In \cite{kallosh} it was shown that this is the most general form of the $Z_2$ action consistent with the symplectic properties of the gravitino. 
The $Z_2$ action on the supersymmetry generating parameter $\epsilon$ must be the same as that on the 4d components of the gravitino.  
 
Further we need to define the $Z_2$ symmetry under reflection around the second fixed point at $x^5=\pi\rho$      
\bea 
\label{z2f} 
\psi^A_\mu(\pi\rho-x^5)= \alpha \gamma_5 \cq^A_B \psi_\mu^B(\pi\rho+x^5) & \psi^A_5(\pi\rho-x^5)= -\alpha\gamma_5 \cq^A_B \psi_5^B(\pi\rho+x^5). 
\eea   
Apart from the conventional case $\alpha=1$, in this letter we also consider the `flipped' supersymmetry  with $\alpha=-1$.  In the latter case supersymmetry is always broken globally, as different spinors survive the orbifold projection on each wall. Note also, that in the flipped case we have $\psi_\alpha^A(x^5+2\pi\rho)=-\psi_\alpha^A(x^5)$. 
 
It is straightforward to check that the  5d {\it ungauged}
supergravity action  is invariant under transformations (\ref{z2}) and
(\ref{z2f})  but the {\it gauged} supergravity action  is not invariant if the prepotential $\cp$ is general.  In reference \cite{kallosh} the constraint stating that  $\cp$ must either commute or anticommute with $\cq$ was derived. We find  that there exists a less restrictive  possibility. The action and the supersymmetry transformation laws are still $Z_2$ invariant if we choose the prepotential of the form: 
\beq  
\label{ourpre} 
g\cp = g_1 \epsilon(x^5) \cR +  g_2 \cS 
\eeq   
where $\cR=\vec{R} \cdot i\vec{\sigma}$ commutes with $\cq$ and $\cS=\vec{S} \cdot i\vec{\sigma}$ anticommutes with $\cq$. Equivalently,  $\cR= i \sqrt{\vec{R}^2} \cq$ and  $\cS = (\vec{Q}\times \vec{U}) \cdot i\vec{\sigma}$ with some arbitrary vector $\vec{U}$. Note that the cosmological constant  does not contain the step function and is given by 
\begin{equation} 
\Lambda_5=-\frac{16}{3}(g_1^2\vec{R}^2 + g_2^2\vec{S}^2). 
\end{equation} 
The supergravity action with prepotential (\ref{ourpre}) contains both symmetric  and antisymmetric (multiplied by $\epsilon(x^5)$) gravitino masses.  
The presence of the $Z_2$-symmetric piece $\cS$ in the prepotential results in 
the supersymmetric detuning between the brane tensions and the bulk cosmological term. 
%The, thus  
%it interpolates between the scenarios of \cite{bagger} and \cite{gp,flp}. 
The part of the gravitino transformation law due to gauging  is now: 
\beq 
\label{susygrav1} 
\delta \psi_\alpha^A = 
-i \frac{\r2}{3}\gamma_\alpha (g_1\epsilon(x^5)\cR^A_B+g_2 \cS^A_B)\epsilon^B.   \eeq   
The presence of the step function in the above transformation law
implies that the 5d action is not supersymmetric. The fifth derivative
in the gravitino kinetic term acts on the step function producing an
expression multiplied by a  delta function. The uncancelled variation is: 
\bea 
\label{varo} 
\delta \cl = -2i\sqrt{2}g_1(\delta(x^5)-\delta(x^5-\pi\rho)) e_4 \cR^A_B \ov{\psi_\mu}_A \gamma^\mu \gamma^5 \epsilon^{B}.  
\eea   
Notice that when $g_1=0$ the above variation  vanishes
implying that the Lagrangian is supersymmetric. 
Using the fact that  the matrix  $\cR$ is proportional to $\cq$ we have the following relations: 
\bea 
\gamma_5\cR^A_B \epsilon^B(0)= i\sqrt{\vec{R}^2}\epsilon^A(0) 
\nn 
\gamma_5\cR^A_B \epsilon^B(\pi\rho)= i\alpha\sqrt{\vec{R}^2}\epsilon^A(\pi\rho). 
\eea 
Thus:  
\bea 
\label{var} 
\delta \cl = 2\sqrt{2}g_1\sqrt{\vec{R}^2} e_4\ov{\psi_\mu}_A \gamma^\mu \epsilon^{A} (\delta(x^5)-\alpha\delta(x^5-\pi\rho)). 
\eea    
The variation (\ref{var}) can be cancelled by the variation of the determinant in the brane tension term: 
\beq 
\cl_T= -4 \sqrt{2}g_1\sqrt{\vec{R}^2} e_4 (\delta(x^5)-\alpha\delta (x^5-\pi\rho)). 
\eeq 
Summarizing (and changing the normalization to that used by Randall and Sundrum), we constructed a {\em locally  supersymmetric Lagrangian}, which has the following bosonic gravity part: 
\beq 
M^{-3} S= \int d^5 x \sqrt{-g_5} (\frac{1}{2}R + 6 k^2)-  6 \int d^5 x\sqrt{-g_4}k T (\delta(x^5) - \alpha \delta(x^5-\pi\rho)) 
\eeq 
where we have defined  
\begin{equation} 
k = \sqrt{\frac{8}{9}(g_1^2 R^2+ g_2^2S^2)} 
\end{equation} 
 and  
\begin{equation} 
T= \frac{g_1\sqrt{\vec{R}^2}}{\sqrt{(g_1^2 R^2+ g_2^2S^2)}}. 
\end{equation} 
  
The BPS relation between the bulk cosmological constant and the brane
tensions (`the Randall-Sundrum fine-tuning') corresponds to $T=1$,
which holds only when $\langle \cS \rangle=0$. In such case the vacuum
solution is  $AdS_5$ in the bulk  with flat Minkowski branes
\cite{rs1}. This vacuum preserves one half of the supercharges
corresponding to unbroken N=1 supersymmetry in four dimensions
\cite{flp}. As soon as we switch on non-zero $\cS$, we get $T < 1$,
the BPS relation  is destroyed and the vacuum breaks
%either $3/4$ of the supersymmetries, corresponding to a $N=1/2$ 
%supersymmetry left in four dimensions.
all supersymmetries. 
This is not surprising, since only the case $\langle
\cS\rangle =0$ (with the `odd gauge charge' and the `antisymmetric
gravitino masses') corresponds to a thin wall limit of smooth $1/2$
BPS supersymmetric domain wall solutions \cite{cvetic}.   

The $N=1$ supersymmetry is broken when $\alpha=\pm 1$ and $\langle \cS \rangle
\neq 0$.  
If $\alpha =-1$, then supersymmetry is always broken globally, 
independently of the expectation value of $\cS$. To see this explicitly, 
let us take $\langle \cS \rangle = 0$ and 
note that the fermions which are allowed to propagate on the left  
and right branes have to obey the conditions $W^{A}_{0\;B} \psi^B=0$ and  
$W^{A}_{\pi \;B} \psi^B=0$ respectively, where $W_{0,\;\pi}$ are given by  
(\ref{z2f}). The projection operators $\Pi^{A}_{\pm \;B} = \frac{1}{2}  
( {\bf 1}  \delta^{A}_B \pm \gamma_5 \cq^{A}_B )$ split each spinor  
into two components, one of which is annihilated by $W_0$:  
$W_0 \epsilon_{+} = W_0 \Pi_{+} \epsilon = 0$. The second component, $\Pi_{-}  
\epsilon$,  
is annihilated by $W_\pi$ if $\alpha=-1$; for $\alpha=+1$ $W_0 = W_\pi$.   
The BPS conditions imply (we take here $ds^2 = a^2 (x^5) dx^2 + (d x^5)^2 $) 
\beq \label{bps3} \frac{a'}{a} \gamma_5 \epsilon^A + \frac{2 \sqrt{2}}{3} g_1 \epsilon(x^5)   
\sqrt{R^2} \cq^{A}_B \epsilon^B = 0 \eeq 
(this holds for $AdS_4$ and Minkowski foliations).
When we apply the operator $\Pi_{+}$ to (\ref{bps3}), we obtain the conditions 
$\frac{a'}{a} + \frac{2 \sqrt{2}}{3} g_1 \epsilon(x^5)   
\sqrt{R^2}=0$ or $\epsilon_{+} \equiv 0$. The first possibility  
leads to discontinuities of the warp factor at the fixed points: 
$ [ \frac{a'}{a} ]_0 = - 2 k T$, $[ \frac{a'}{a} ]_{\pi \rho} = + 2 k T$. 
However, the matching conditions in the equations of motion give  
$ [ \frac{a'}{a} ]_0 = - 2 k T$, $[ \frac{a'}{a} ]_{\pi \rho} = + 2  
\alpha k T$, which are in contradiction with the BPS condition for  
$\alpha =-1$, unless $\epsilon_{+} \equiv 0$. Applying to (\ref{bps3}) the  
second projector, $\Pi_{-}$, one finds out immediately that boundary 
conditions and BPS conditions agree on both branes only 
for $\epsilon_{-} \equiv 0$.  
Thus there exists no globally defined Killing spinor in the setup with  
flipped $Z_2$ acting on fermions (bosons are acted on as in the unflipped case), and all supersymmetries are broken, in particular the 4d effective theory  
is explicitly nonsupersymmetric.  
 
\section{Non-BPS Branes} 
   
In the remainder of this letter we construct static vacuum solutions in the detuned case ($T < 1$ and/or $\alpha=-1$).       
We start with the case $\alpha=1$, i.e. opposite brane tensions. In \cite{bd,bin}  a method applicable to  branes violating the BPS relation was outlined. The idea was to write the BPS solution in time-boosted coordinates, so that it satisfied the matching conditions for $T>1$ .       
Here, we perform a similar trick which works for $T<1$.    
 
We write the  bulk solution in the conformal gauge  
\beq 
\label{bpss} 
ds^2= a_{BPS}^2(y)(\eta_{\mu\nu}dx^\mu dx^\nu + dy^2). 
\eeq  
In the case at hand  
\begin{equation} 
a_{BPS}(y)=1/(k y). 
\end{equation} 
Then we apply  a rotation to the bulk defining the  new coordinates  parametrized by the three-vector  $\vec{h}$ ($i=1,2,3$) 
\bea 
&&\tilde{x^i} = \frac{x^i - h^i y}{\sqrt{1+h^2}} 
\nn\\ 
&&\tilde{y} = \frac{y+ h^i x^i}{\sqrt{1+h^2}} 
\eea   
and $\tilde x_0=x_0$, with $h$ being the norm of $\vec{h}$.   
Under this change of coordinates $\tilde{g}_{\alpha\beta}(\tilde{x})=g_{\alpha\beta}(x)$ so we can rewrite (\ref{bpss}) as: 
\beq \label{eq20}
ds^2=
a_{BPS}^2(\frac{\tilde{y}-h^i\tilde{x^i}}{\sqrt{1+h^2}})(\eta_{\mu\nu}d\tilde{x}^\mu
d\tilde{x}^\nu + d\tilde{y}^2). 
\label{solut}
\eeq  
This is the same solution as  (\ref{bpss}) written in another
coordinate system so this is also a legitimate solution in the bulk. 
Let us check that this provides a solution in the bulk which satisfies
the marching condition
\begin{equation} 
\frac{\pa_{\tilde{y}} a}{{a^2}}=- k T. 
\end{equation} 
for a scale factor $a$.
Note that  
\begin{equation} 
\frac{\pa_{\tilde{y}} a_{BPS}}{{a^2_{BPS}}} = - k(1+h^2)^{-1/2}, 
\end{equation} 
hence we find  that (\ref{solut}) leads to a solution provided 
\begin{equation} 
\sqrt{1+h^2} T=1.  
\end{equation} 
Thus, we have to perform a rotation parametrized by a vector $\vec{h}$ whose norm is  
\beq \label{eq24}
h = \frac{\sqrt{1-T^2}} {T}.  
\eeq 
Alternatively, one can view this solution as  $AdS_5$ bulk with  'tilted' branes, i.e. the transverse position of the branes depends on $x^\mu$ and is given by  
\begin{equation} \label{secb}
y_1= -h_{1}^i x_{1}^i+\sqrt{1+h^2} \;\pi\rho 
\end{equation} 
for the second brane.
To match boundary conditions on the first brane one can imagine playing the same trick, i.e. rotating the brane by a vector $h_{0}^i,\;\;i=1,2,3$, independent
of $\vec{h}_1$. Again, locally near the brane the bulk solution is $AdS_5$ 
in the  conformal frame, and the equation defining the brane is 
\beq \label{fstb}
y_0 = - h_{0}^i x_{0}^i .
\eeq
Since in the bulk the two local $AdS_5$ solutions match smoothly onto each other, the branes given by (\ref{secb}) and (\ref{fstb}) plus the $AdS_5$ metric in the bulk (in conformal coordinates) form a complete solution to the equations of motion. 

 The breaking of supersymmetry can be analysed along the lines
of the previous section where Killing spinors were discussed in the
flipped case, using the metric (\ref{eq20}). 
One finds that the tilted configuration breaks
supersymmetry completely.

Let us specialize to the case where both vectors point in the $x_3$
direction. The branes are rotated with respect to the $x_3$--axis  by angles given by
\beq
\tan \theta_1 = - h^{3}_1, \;\;\; \tan \theta_0 = - h^{3}_0 .
\eeq
Since $h^{3}_0=\pm h^{3}_1$ one obtains $\tan \theta_k = \pm \frac{\sqrt{1-T^2}}{T}$, $k=0,1$. It follows that the rotation angles of the tilted branes 
are proportional to the supersymmetry breaking expectation value of  the 
$Z_2$--even part of the gravitino supersymmetry transformations, $\langle \cS \rangle$,
\beq
\tan \theta_k = \pm \frac{g_2 \sqrt{S^2}}{g_1 \sqrt{R^2}}.  
\eeq
One can formulate this observation in a different way: the rotation of a brane 
induces $N=1$ supersymmetry breakdown whose magnitude is 
proportional to the rotation angle. 
The second brane also needs to be
rotated to match the boundary conditions properly.

\vskip0.2cm 
Now we move on to the `flipped susy'  case $\alpha=-1$. A vacuum solution  in the warped product form can be found 
\begin{equation} 
ds^2 = a^2(x^5)g_{\mu\nu}dx^\mu dx^\nu + R_0^2 (dx^5)^2, 
\end{equation} 
where $g_{\mu\nu}$ is the $AdS_4$ metric with cosmological constant $\bar \Lambda$ 
\beq 
g_{\mu\nu}dx^{\mu}dx^{\nu}=e^{-2\sqrt{-\bar\Lambda}x_3}(-dt^2+dx_1^2+dx_2^2) +dx_3^2 
\eeq  
and the third coordinate $x_3$ has been singled out. 
As long as $T<1$  the static vacuum solution is $AdS_5$ in the bulk  and  the warp factor  can be parametrized as \cite{dewolfe,rk}: 
\beq 
\label{ads4sol} 
a(x^5) = \frac{\sqrt{-\bar \Lambda}}{k}\cosh (k R_0 |x_5| - C). 
\eeq 
This metric is isometric to the warped metric of $AdS_5$:  
\begin{equation} 
ds^2=e^{-2 k R_0 y}(-d\tilde t^2+d\tilde x_1^2+d\tilde x_2^2+ d\tilde x_3^2)+R_0^2dy^2, 
\end{equation} 
where 
\begin{equation} \label{txt}
\tilde x_3= (\sqrt{-\bar\Lambda})^{-1}\hbox{tanh}(C-kR_0x_5)\;e^{\sqrt{-\bar \Lambda}x_3},\ y=\sqrt{-\bar\Lambda}\frac{x_3}{k R_0}-\frac{\ln(\frac{\sqrt{-\bar\Lambda}}{k} \hbox{cosh}(C-k R_0 x_5))}{k R_0} 
\label{cor} 
\end{equation} 
and $\tilde t=t,\ \tilde x_1=x_1, \ \tilde x_2=x_2$. 
This coordinate transformation will be useful in the following section where 
we discuss the existence of a geodesically complete submanifold of $AdS_5$ with a horizon where the two branes can be embedded. 
 
The matching  conditions  for $AdS_4$ branes embedded in $AdS_5$ read 
\bea 
\tanh(C)&=&  T \\  
\tanh (k R_0 \pi\rho - C)&=& -\alpha  T\; = \;  T .  
\eea 
The first condition sets the integration constant $C$ and the second fixes the size of the fifth dimension.  
The radion is stabilized at the value 
\begin{equation} \label{radfix}
\pi\rho k R_0= \ln(\frac{1+T}{1-T}). 
\end{equation} 
Moreover, the magnitude of the brane cosmological constant is fixed by 
the normalization $a(0)=1$. This leads to 
\begin{equation} 
\bar \Lambda=(T^2 - 1)k^2 < 0. 
\end{equation} 
The cosmological constant on the brane depends directly on the scale of supersymmetry breaking on the brane. The same is true for 
the expectation value of the radion. Notice, that $\langle R_0 \rangle$ 
can be expressed solely in terms of $k$ and $\bar \Lambda$. 
The formula equivalent to (\ref{radfix}) in the case $\alpha=+1$  
gives $R_0 = 0$.  
To summarize, the nonzero expectation value of $\cS$ gives rise to detuning
between brane and bulk tensions and as a consequence to stabilization of 
the radion. The $\langle \cS \rangle $ contributes also to supersymmetry breakdown, but in the case of $\alpha=-1$ supersymmetry is broken even if 
$\langle \cS \rangle = 0$.

One finds that (\ref{ads4sol}) is not a valid solution in the case $T=1$. 
Indeed, this implies that the brane cosmological 
constant vanishes and the second brane is sent to infinity. 
In that case we expect a global mismatch due to the boundary 
condition on one of  
the branes. This is the subject of the following section.     
 
\section{Flipped Supersymmetry and Moving Branes} 
 
In this section we deal with the flipped case $\alpha=-1$ 
corresponding to opposite chiralities of the boundary branes and focus on the case where  
the brane tensions on both branes are equal. 
In order 
to find non-singular solutions to the equations of motion it is 
sufficient to use locally AdS spaces in the bulk. 
Consider the following bulk metric with $l=1/k$ 
\begin{equation} \label{locmet}
ds^2= -(\frac{r^2}{l^2}-1)dt^2 +\frac{1}{\frac{r^2}{l^2}-1}dr^2 +r^2d\Sigma^2 
\end{equation} 
where $d\Sigma^2$ 
is the metric on a compact hyperbolic three-space $\Sigma$  of constant 
curvature $-1$. Such spaces can be obtained by quotienting the 
usual hyperbolic three-space by a discrete group. The metric is 
the usual AdS-black hole metric with a hyperbolic horizon where 
the mass of the black hole has been set to zero. Locally one can define, 
\cite{emparan},
\begin{equation} \label{locods}
u=-\xi\sqrt{1-\frac{l^2}{ r^2}}e^{-t/l},\ v=\xi\sqrt{1-\frac{l^2}{r^2}}e^{t/l} 
\end{equation} 
and 
\begin{equation} 
\tilde r= l\frac{r}{\xi}, 
\end{equation} 
where the metric on the hyperbolic three-space is 
\begin{equation} 
d\Sigma^2=\frac{d\xi^2+dx_idx^i}{\xi^2}. 
\end{equation} 
This leads to the  $AdS_5$ metric in Poincare coordinates 
\begin{equation} \label{pocods}
ds^2=\frac{\tilde r^2}{l^2}(-dudv+d\tilde x_id\tilde x^i)+\frac{l^2}{\tilde r^2}d\tilde r^2. 
\end{equation} 
We have used the obvious relation between the light cone coordinates and orthogonal 
coordinates   
\begin{equation} 
u=\tilde t-\tilde x_3\  v=\tilde t+\tilde x_3 
\end{equation} 
and identified $\tilde x_i=x_i$. 
Globally one must restrict the range of $r$ to $r\ge l$ where the horizon 
sits at $r=l$. The 
Carter-Penrose diagram can be seen in figure 1 where the maximal 
analytic extension of the manifold ${\cal M}$ (defined by (\ref{locmet})--(\ref{pocods})) has been drawn. It consists of 
two copies of the same $r\ge l$ space glued at $r=l$ by a wormhole 
called the Einstein-Rosen bridge whose topology is the one of $\Sigma$. 
Notice that the horizon sits in the middle of the Carter-Penrose diagram of $AdS_5$. 
 
\begin{figure} 
\epsfxsize=6.cm 
$$\epsfbox{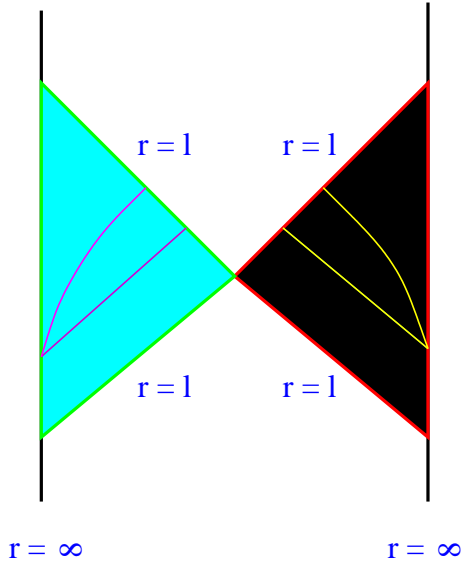}$$ 
\caption{The Carter-Penrose diagram of the geodesically complete manifold 
${\cal M}$ embedded in the $AdS_5$ strip. Both sides of the horizon $r=l$ correspond to $r\ge l$. Two light cones  have been depicted. The moving branes 
stay within the respective light cones reaching the speed of light at  the horizon.} 
\end{figure}

Let us come back to the flipped supersymmetric brane model where 
two branes of the same tension are present.  
It is relevant to locate the two branes with respect to the horizon. 
Notice that the parametrization (\ref{locods}) of ${\cal M}$ implies 
that $\tilde x_3\ge 0$.  
However, the two $AdS_4$ branes of the previous section are such that $\hbox{tanh}(kR_0x_5-C)$ 
have opposite signs. Given that we are interested in solutions related to 
the static  
$AdS_4$ solutions, this leads to the assignment of the opposite sign $\tilde x_3$ coordinates for the two branes in the figure 1. 
Indeed, the  static branes are located at  
\begin{equation} 
x_5=0,\ x_5=\frac{2C}{kR_0}. 
\end{equation} 
The horizon at $r=l$ corresponds to  $\tilde x_3=0$ and that through (\ref{txt}) 
corresponds to $
%\begin{equation} 
x_5=\frac{C}{kR_0}, 
%\end{equation} 
$
the point lying just in the middle between the static branes.  
We thus find that the two branes are on both sides of the horizon.  
As $T\to 1$ the second brane goes to infinity and the static solution is not valid anymore. We will find that, by embedding the two branes in ${\cal M}$,  this problem can be alleviated.

One can find a 
solution to this model by embedding each brane in one of the two copies 
of the local AdS space-time $r\ge l$. The two branes are 
separated by the hyperbolic Einstein-Rosen bridge located at the horizon. 
In each half space the motion of the branes is determined by the 
the boundary condition for moving branes 
\begin{equation} 
(\frac{r^2}{l^2}-1 +\dot r^2)^{1/2}=T\frac{r}{l} 
\end{equation} 
in proper time $\tau$ defined by 
\begin{equation} 
-(\frac{r^2}{l^2}-1)\dot t^2  +\frac{1}{\frac{r^2}{l^2}-1}\dot r^2=-1. 
\end{equation}  
The boundary condition is equivalent to the Friedmann equation 
\begin{equation} 
H^2=\frac{T^2-1}{l^2}+\frac{1}{r^2} 
\end{equation} 
where $H=\dot r /r$. 
Now the induced metric on the brane reads 
\begin{equation} 
ds_B^2=-d\tau^2 + r^2(\tau)d\Sigma^2. 
\end{equation} 
The solution to Friedmann's equation is 
\begin{equation} 
r(\tau)=\pm\frac{l}{\sqrt{1-T^2}}\sin( \frac{\sqrt{1-T^2}}{l}\tau)  
\end{equation} 
leading to the identification of the moving brane with an $AdS_4$ brane with cosmological constant 
$\bar \Lambda$.  
   Notice that $r\ge l$ implies that $\tau$ does not cover the whole circle. 
This implies that the moving branes are only a slice of $AdS_4$ branes, 
with an infinite period  oscillation in the $t$ coordinate. 
In conclusion the moving brane is nothing but another way of embedding $AdS_4$ branes in $AdS_5$ which  
is only locally isometric to the static branes of section 3.

We can now treat the degenerate case $T =1$ 
for which the above brane cosmological constant $\bar{\Lambda} \sim
(T^2-1)$
vanishes.
%for which the cosmological constant vanishes. 
Now the brane has the topology of an open FRW space-time with curvature $k=-1$.  
The motion is given by 
\begin{equation} 
r(\tau)=\pm \tau +r_0.  
\end{equation} 
We can now deduce the evolution equation 
\begin{equation} 
\frac{r^2(t)}{l^2}-1=e^{\pm2t/l}. 
\end{equation} 
This implies that the brane either recedes  away from the horizon or 
converges towards the horizon. Moreover it takes an infinite amount of time 
to reach the horizon. This is to be contrasted with the finite amount of proper time 
that it takes to reach the horizon.

It convenient to use tortoise coordinates defined by 
\begin{equation} 
u=\frac{l}{2}\ln(\frac{\frac{r}{l}-1}{\frac{r}{l}+1}) 
\end{equation} 
leading to 
\begin{equation} 
ds^2=-(\hbox{cotanh} ^2\frac{u}{l} -1)(dt^2-du^2)+l^2 \hbox{cotanh} ^2 
\frac{u}{l} d\Sigma^2. 
\end{equation} 
The horizon is now at $u=-\infty$ and the spatial infinity at $u=0$. 
In the vicinity of the horizon the metric becomes 
\begin{equation} 
ds^2=4e^{2u/l}(-dt^2+du^2) + l^2 d\Sigma^2. 
\end{equation} 
Putting $x=2le^{u/l}$ we find that the horizon has the 
structure of a Rindler space times the hyperbolic space $\Sigma$. 
In particular the temperature of the horizon is  
\begin{equation} 
T_H=\frac{1 }{2 \pi l}.  
\end{equation} 
It can be interpreted as the emission of gravitons and gravitini coming 
out through the wormhole.  
One may wonder 
how an observer bound to a brane would interpret the supersymmetry breakdown 
seen on his/her brane, with no  classical communication with the second
brane from 
which he/she is separated by the horizon. We note  
that there is a nonzero temperature of the horizon, of  order  $k$, 
which is observable, and can be seen locally on a given side of the
horizon as the reason for the mass splittings in the bulk multiplets.  
One could describe this phenomenon as temperature 
mediated supersymmetry breakdown.

In the system of coordinates that we adopted the light-like geodesics follow 
\begin{equation} 
u=r_0 \pm t 
\end{equation} 
showing that particles take an infinite amount of time to reach the horizon. 
Notice that the trajectory of the moving brane is given by 
\begin{equation} 
u(t)=\frac{l}{2}\ln(\frac{\frac{r(t)}{l}-1}{\frac{r(t)}{l}+1}) 
\end{equation}  
corresponding to a speed 
\begin{equation} 
\dot u=\pm \frac{l}{r}. 
\end{equation} 
The brane reaches the speed of light at the horizon. 

Let us make a comment on trapping of gravity on the brane.
When the brane is far away from the horizon, $r(t) \gg l$, the metric 
(\ref{locmet}) takes the form of the $AdS_5$ metric in conformal gauge, 
and the brane is moving very slowly with respect to it. Hence, locally
around the brane gravity should be indistinguishable from that around the positive tension Randall--Sundrum brane. However, when the brane comes close
to the horizon, the specific shape of the metric near horizon 
becomes important, and also the Hubble length on the brane, $d_H=H^{-1}=r(t)$,
 becomes comparable to $l$, hence gravity on the brane becomes nonstandard. 
  
\section{Summary}

In this paper we have presented generalized supersymmetric brane worlds. 
We have enhanced supersymmetry transformations by including both 
$Z_2$--odd and $Z_2$--even components in the 5d prepotential. Further,
we have allowed the $Z_2$ symmetry of the Lagrangian to act on fermions
locally, differently in the vicinity of each wall (this is the case of 
flipped boundary conditions for fermionic degrees of freedom). 
In models with flipped boundary conditions or with nonzero vacuum 
expectation value for $Z_2$--even parts of supersymmetry transformations 
($\sim \langle \cS \rangle$) 
the $N=2$ supersymmetry is broken down to nothing by vacuum configurations. 
These configurations, which are non--BPS domain walls in 5d, have the form 
of rotated branes if $\langle \cS \rangle \neq 0$, with supersymmetry breakdown 
proportional to the rotation angle. There are no singularities, but 
4d Lorentz invariance is broken in these vacua. 

Models with flipped boundary conditions correspond to $Dp$--$D\bar{p}$ 
brane--antibrane systems in 5d. There supersymmetry is broken, 
which can be understood as a mismatch between possible Killing spinors 
in the bulk and boundary conditions on the branes. There exist static 
$AdS_4$ branes with a stabilized radion.  
Other non--BPS branes 
that we 
have found are moving, and in addition there is a horizon 
(but no essential singularity) separating the branes. One may wonder 
how an observer bound to a brane would interpret supersymmetry breakdown 
seen on his brane, without classical communication with the second brane. 
The relevant observation is that there is a nonzero temperature of the
horizon, of the order of $k$, which is observable, and can be seen
locally on a given side of the horizon as the reason for the mass splittings in the bulk multiplets. 
One could describe this phenomenon as temperature 
mediated supersymmetry breakdown. An interesting problem is to add brane matter to the backgrounds we have discussed, and to derive low--energy phenomenology of such models. We leave this issue to a future publication.   \\
The observation that in the flipped case one does not  find a background 
solution in the form of a Minkowski foliation is a hint that the usual 
stability and cosmological constant problems are likely to appear in 
locally supersymmetric `flipped' model building.

\vskip0.7cm
The work of A.F. and Z.L. has been supported by RTN programs HPRN-CT-2000-00152
 (Z.L.) and HPRN-CT-2000-00148 (A.F.), 
and by the Polish Committee for Scientific Research grant
5~P03B~119 20 (2001-2002).
 
\end{document}